\documentclass[twocolumn, prl, 10pt, superscriptaddress]{revtex4-1}
\usepackage{graphicx, subfigure, amsfonts,amssymb,amsmath, array, gensymb,fontenc,color, lipsum}

\begin{document}

\title{Electrically Induced Dirac Fermions in Graphene Nanoribbons}

\author{Michele.~Pizzochero}
\email{mpizzochero@g.harvard.edu}
\affiliation{School of Engineering and Applied Sciences, Harvard University, Cambridge, Massachusetts 02138, United States}

\author{Nikita~V.~Tepliakov}
\affiliation{Departments of Materials and Physics, Imperial College London, London SW7 2AZ, United Kingdom}
\affiliation{The Thomas Young Centre for Theory and Simulation of Materials,Imperial College London, London SW7 2AZ, United Kingdom}
\affiliation{Center for Information Optical Technologies, ITMO University, Saint Petersburg 197101, Russia}

\author{Arash~A.~Mostofi}
\affiliation{Departments of Materials and Physics, Imperial College London, London SW7 2AZ, United Kingdom}
\affiliation{The Thomas Young Centre for Theory and Simulation of Materials, Imperial College London, London SW7 2AZ, United Kingdom}

\author{Efthimios~Kaxiras}
\affiliation{School of Engineering and Applied Sciences, Harvard University,  Cambridge, Massachusetts 02138, United States}
\affiliation{Department of Physics, Harvard University, Cambridge, Massachusetts 02138, United States}

\begin{abstract}
Graphene  nanoribbons  are  widely  regarded  as  promising  building  blocks  for  next-generation carbon-based  devices. A critical issue to their prospective applications is whether  and  to  what  degree  their  electronic  structure  can  be externally controlled. Here, we combine simple model Hamiltonians with extensive first-principles calculations to investigate the response of armchair graphene nanoribbons to transverse electric fields. Such fields can be achieved either upon laterally gating the nanoribbon or incorporating ambipolar chemical co-dopants along the edges. We  reveal  that the  field  induces  a  semiconductor-to-semimetal  transition,  with  the  semimetallic phase featuring zero-energy Dirac fermions that propagate along the armchair edges.  The transition occurs at critical fields that scale inversely with the width of the nanoribbons. These findings are universal to group-IV honeycomb  lattices,  including  silicene  and  germanene  nanoribbons, irrespective of the type of edge termination. Overall,  our  results  create  new opportunities  to  electrically  engineer  Dirac  fermions  in  otherwise  semiconducting  graphene-like nanoribbons.
\end{abstract}
\maketitle

\paragraph{Introduction.} Graphene nanoribbons (GNRs) --- few-nanometer wide strips of hexagonally bonded carbon atoms --- are a prospective platform for future nanoelectronic devices owing to their particular combination of quantum confinement effects and  $\pi$-conjugation \cite{Yazyev2013, Zongping2020}. Progress in fabrication techniques has unlocked the possibility of synthesizing GNRs in an atom-by-atom fashion \cite{Cai2010a, Yano2020, Zongping2020}, hence leading to nanoribbons with diverse edge geometries \cite{Cai2010a, Ruffieux2016, Nguyen2017, Liu2015}  and widths \cite{Chen2015a, Jacobse2017, Wang2017, Pizzochero2020a} that can exhibit an array of novel physical phenomena including metal-free magnetism \cite{Li2019, Lawrence2020, Friedrich2020, Sun2020a, Pizzochero2021} and topological bands \cite{Rizzo2018, Sun2020b, Groning2018, Cao2017}.

Within the variety of GNRs obtained so far, armchair graphene nanoribbons (AGNRs) are arguably the best suited for device integration \cite{Zongping2020, Jacobse2017} by virtue of their favorable fabrication scalability \cite{DiGiovannantonio2018}, facile transferability onto target substrates \cite{Bennett2013, BorinBarin2019}, and long-term chemical stability under ambient conditions \cite{BorinBarin2019}. Unlike zigzag graphene nanoribbons, AGNRs are routinely integrated as active channels in field-effect transistors operating at room temperature \cite{Li2008, Martini2019} that often display superior performance among graphene-based devices \cite{Llinas2017}. The atomic-scale control achieved in fabrication of AGNRs has enabled the realization of complex one-dimensional nanoarchitectures, including two- \cite{Blankenburg2012, Ma2019}, multi- \cite{Cai2010a}, and hetero-terminal junctions \cite{Bronner2018, Cai2014}, which are important steps toward all-carbon nanocircuitry \cite{Jacobse2017, Kang2013, Martini2019}. The logic capabilities of AGNRs stem from their width-modulated energy gaps, the magnitude of which scales inversely with the number of carbon atoms across the nanoribbon axis, $n$ \cite{Li2008, Chen2013}. Specifically, the energy gaps decay with $n$ according to multiple-of-three oscillations, depending on whether $n = 3p$, $3p+1$, or $3p+2$, with $p$ being a positive integer \cite{Son2006a, Yazyev2013}. 

Devising strategies to manipulate the electronic structure of graphene nanoribbons via external stimuli is of crucial importance to the development of carbon-based electronics. Earlier theoretical works predicted that transverse electric fields induce half-metallic phases and spin-polarized currents in zigzag graphene nanoribbons \cite{Son2006b, Kan2007, Wang2012, Kan2008}. However, analogous field effects in armchair graphene nanoribbons remain poorly understood \cite{Novikov2007, Raza2008, Roslyak2010}. In this Letter, we show that transverse electric fields enforce a semiconductor-to-semimetal transition in AGNRs, with the gapless phase possessing zero-energy Dirac fermions that propagate along the nanoribbon edges. The electric field can be applied externally, \emph{i.e.}, by laterally contacting the nanoribbons to a pair of electrodes, or internally, \emph{i.e.}, by ambipolar chemical co-doping of the edges. This width-dependent transition is inherent to group-IV nanoribbons, including recently realized silicene and germanene nanoribbons, irrespective of their edge passivation.  Our findings pave the way to achieve and control Dirac fermions in the ultimate limit of miniaturization.

\begin{figure}[t]
    \centering
    \includegraphics[width=1\columnwidth]{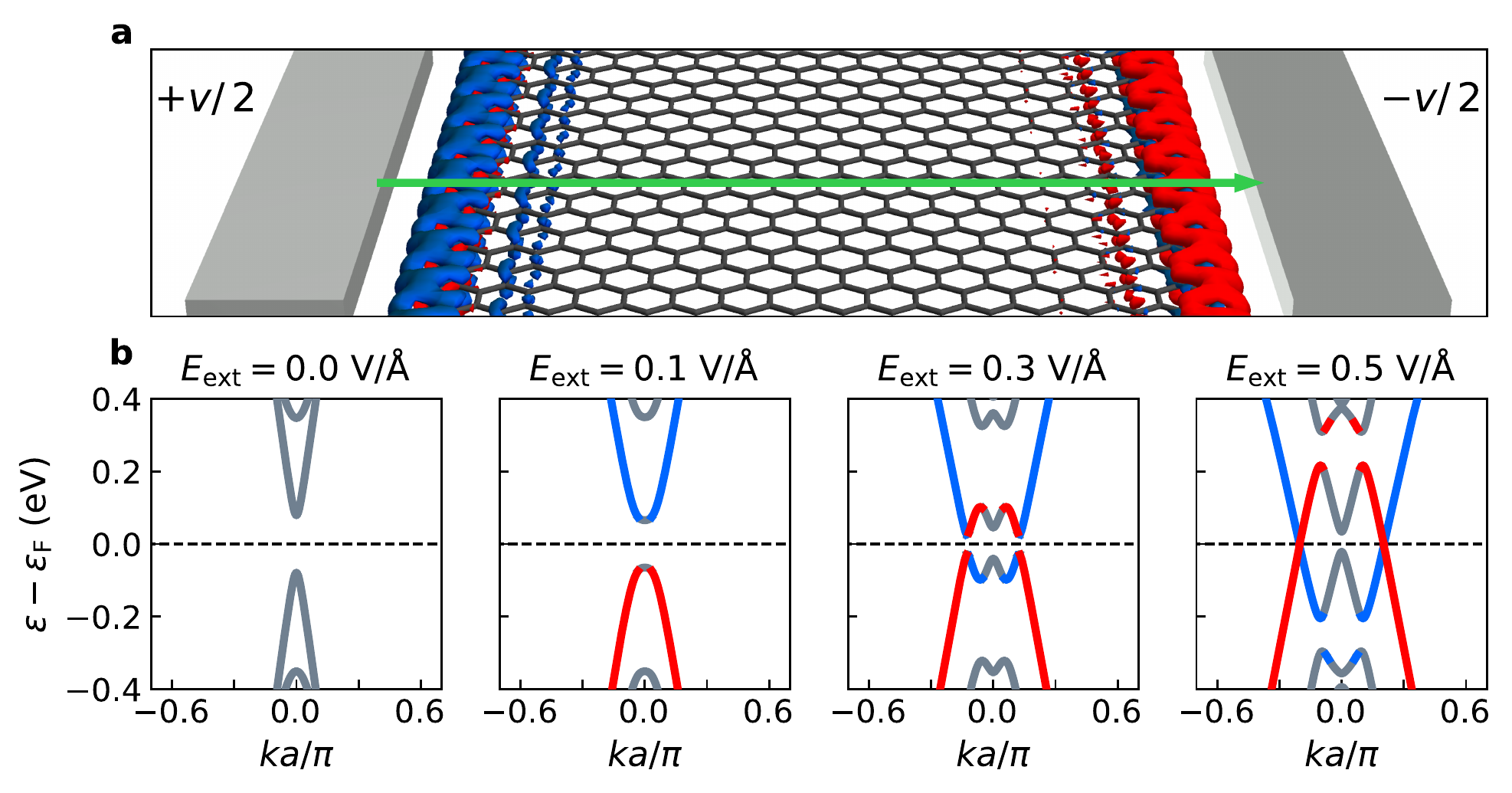}
    \caption{\textbf{Dirac fermions in graphene nanoribbons induced by external electric fields.} (a) Atomic structure of a hydrogen-terminated armchair graphene nanoribbon of width $n = 33$ in a transverse external electric field. Such a field can be experimentally realized by laterally contacting the nanoribbon to source and drain electrodes, as illustrated. Red and blue isosurfaces represent the electron- and hole- charge density induced by the field. Isosurface values are set to $\pm 0.001$ $e$/{\AA}$^3$. (b) Band structures of AGNR at electric fields $E\textsubscript{ext} = 0.0$, $0.1$, $0.3$, $0.5$~V/{\AA}. Red and blue lines indicate the bands that mainly originate from each of the two armchair edges of the nanoribbon. Energies are referenced to the Fermi level, $\varepsilon\textsubscript{F}$. \label{Fig1}}
\end{figure}

\smallskip
\paragraph{Electric route to Dirac fermions in AGNRs.} 
We begin by considering the response of hydrogen-terminated graphene nanoribbons to external electric fields. We perform density-functional theory calculations using the gradient-corrected approximation to the exchange-correlation functional devised by Perdew, Burke, and Ernzerhof \cite{PBE}. We have explicitly verified that our results are insensitive to this particular choice of functional. Additional computational details are provided in Supporting Note 1. We introduce an external electric field, $E\textsubscript{ext}$, in the direction transverse to the axis of the nanoribbon. This situation can be realized experimentally in an open circuit split-gate configuration where the armchair edges of the nanoribbon are contacted to source and drain electrodes \cite{Son2006b}, as schematically illustrated in Fig.~\ref{Fig1}(a). The charge density shown in the same figure indicates that such an electric field drives a spatial separation of electrons and holes along the two edges of the nanoribbon.

In Fig.~\ref{Fig1}(b), we show the evolution of the band structure of a hydrogen-terminated armchair graphene nanoribbon of width $n = 33$ (or, in the unit of length, $w = 3.9$ nm, where $w = (n-1)\sqrt3d\textsubscript{C} / 2$ and $d\textsubscript{C} = 1.42$ {\AA} is the carbon-carbon bond length) as a function of $E\textsubscript{ext}$; results for other widths are given in Supporting Figs.~S1-S3. As the strength of the field increases, the band extrema reorganize by first narrowing and then quenching the direct energy gap, while shifting in momentum from $ka = 0$ to $ka = \pm 0.3 \pi $. At $E\textsubscript{ext} \approx 0.5$~V/{\AA}, valence and conduction bands cross at the Fermi level, giving rise to zero-energy Dirac points. The projections of the band structure onto the carbon atoms that form the two armchair edges reveal that these linearly dispersing states are confined selectively on these edges, as is further confirmed by the local density of states shown in Supporting Fig.~S4. This indicates that Dirac fermions with opposite Fermi velocities propagate along each of the two edges in AGNRs when subjected to a uniform external electric field. This effect is independent of the width of the nanoribbon; see Supporting Fig.~S5.

To unveil the origin of the field-induced crossover from the semiconducting to the Dirac semimetallic phase, we rely on a tight-binding Hamiltonian of the $\pi$-electron network,
\begin{equation}
\mathcal{H} = \sum_i v_i c_i^{\dagger} c_i - \sum_{i,j} t_{ij} (c_i^{\dagger} c_j + \mathrm{h.c.}),
\label{Eqn1}
\end{equation}
where $c_i^{\dagger}$ and  $c_i$ are the fermion creation and annihilation operators, respectively, $v_i$ is the on-site energy acting on the $i$th lattice site, and $t_{ij}$ is the hopping integral between the $i$th and $j$th sites. We include the first, second, and third nearest-neighbors hopping integrals with the respective values of $t_1 = 2.70$ eV,  $t_2 = 0.20$ eV, and  $t_3 = 0.18$ eV \cite{Kundu2011,Pizzochero2020b}, and vary the strength and spatial distribution of $v_i$ to examine its role on the electronic properties.

\begin{figure}[t]
    \centering
    \includegraphics[width=1\columnwidth]{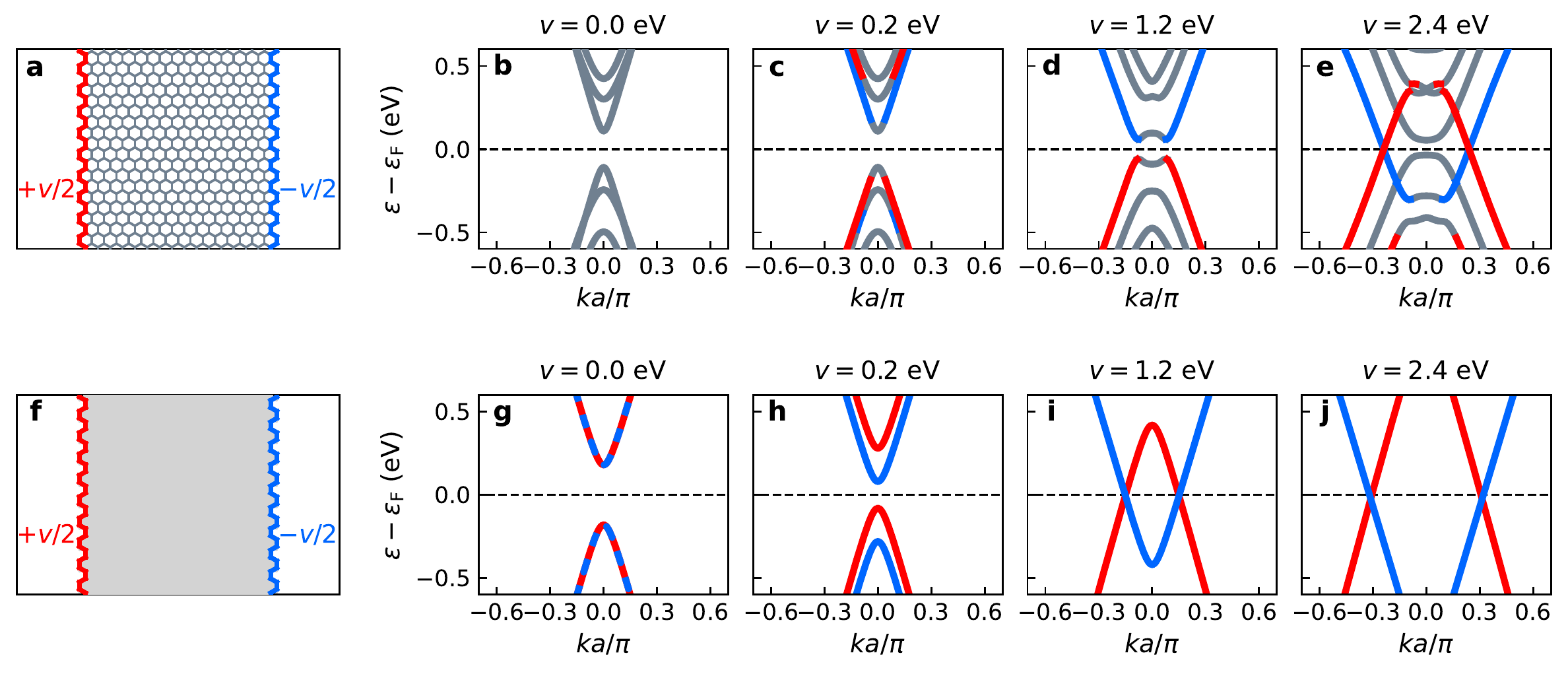}
    \caption{\textbf{A simple model to explain the field-induced Dirac fermions in graphene nanoribbons.} (a) Atomic structure of an AGNR of width $n = 33$, along with (b-e) its band structure with the on-site energies linearly varying from $v/2$ to $-v/2$ between the two edges. (f) Atomic structure of a pair of \emph{cis}-polyacetylene chains, a simple model to describe the edges of an armchair graphene nanoribbon, along with (g-j) its electronic band structure upon the introduction of opposite on-site energies $\pm v/2$ on the two chains. Results obtained from the tight-binding Hamiltonian given in Eqn.~(\ref{Eqn1}). Red and blue lines indicate the bands that mainly originate from each of the two \emph{cis}-polyacetylene chains composing the armchair edges of the nanoribbon. Energies are referenced to the Fermi level, $\varepsilon\textsubscript{F}$.
    \label{Fig2}}
\end{figure}

In Fig.~\ref{Fig2}(a), we consider a representative AGNR of width $n = 33$; results for other widths are given in Supporting Fig.~S6. To emulate the action of the electric field, we set the on-site energies such that they vary linearly from $v/2$ to $-v/2$ between the two armchair edges. The evolution of the band structure with increasing $v$ is given in Fig.~\ref{Fig2}(b-e). As on-site energies increase, the energy gap of the system decreases and eventually vanishes, yielding band crossing at the Fermi level. As evidenced by the projections of the energy bands, on-site energies of opposite signs cause electrons to redistribute from one edge to the other through the inner sites of the nanoribbon, leading to gapless Dirac states.

To further clarify the emergence of these Dirac states, we consider a simple model consisting of a pair of parallel \emph{cis}-polyacetylene chains, as shown in Fig.~\ref{Fig2}(f). This system can be regarded as the simplest description of the edges of AGNRs. %We assume that the electronic subsystems of the two chains do not interact, but electrons can redistribute between the chains if their Fermi levels are different. 
At vanishing on-site energies, the band structure shown in Fig.~\ref{Fig2}(g) consists of doubly degenerate valence and conduction bands that, similar to AGNRs, have an energy gap at the center of the Brillouin zone \cite{Son2006a}. Upon the introduction of $\pm v/2$ on the two chains, the four bands split into two sets, as displayed in Fig.~\ref{Fig2}(h-j). Importantly, the top of the valence band of one chain shifts to the higher energies, while the bottom of the conduction band of the other chain shifts to lower energies. As a result, the energy gap of the system reduces and ultimately closes at large enough values of $v$, with the linearly dispersing bands of the two chains crossing at the Fermi level. Hence, opposite on-site energies induce a semimetallic phase featuring two Dirac points located at equivalent off-symmetry points of the Brillouin zone. Although each \emph{cis}-polyacetylene chain is individually insulating, the inter-chain redistribution of electrons allows the formation of the semimetallic phase. The analogy between Dirac points in the armchair edges of a graphene nanoribbon and \emph{cis}-polyacetylene chains is made clear by the comparison of Fig.~\ref{Fig2}(e) with Fig.~\ref{Fig2}(i). In both cases, the semimetallic states are localized on each of the two armchair chains of carbon atoms. These conceptually simple models illustrate that the onset of the field-induced Dirac semimetallic phase in otherwise semiconducting AGNRs traces back to edges having charges of opposite sign and the resulting band shifts.

\begin{figure}[t]
    \centering
    \includegraphics[width=1\columnwidth]{Fig3.pdf}
    \caption{\textbf{Effect of the width of the nanoribbon on the field-induced Dirac fermions.} (a--c) Energy gaps, $\varepsilon\textsubscript{g}$, of hydrogen-terminated armchair graphene nanoribbons as a function of an increasing electric field, $E\textsubscript{ext}$. Depending on their width, $n$, AGNRs are classified into the (a) $n = 3p$, (b) $n = 3p+1$, and (c) $n = 3p+2$ classes, with $p$ being a positive integer. (d--f) Evolution of the critical field, $E\textsubscript{crit}$, required to close the energy gap as a function of the width of the nanoribbon for the (d) $3p$, (e) $3p+1$, and (f) $3p+2$ classes. Circles represent first-principles results while solid lines are the fit with Eqn.~(\ref{Eqn2}) to extract the class-dependent critical potential, $V\textsubscript{crit}$. \label{Fig3}}
\end{figure}

\smallskip
\paragraph{Role of width and chemical composition.}  We generalize our findings by systematically addressing the dependence of the semiconductor-to-semimetal transition on the width of hydrogen-terminated AGNRs. In Fig.~\ref{Fig3}(a-c), we trace the energy gap, $\varepsilon\textsubscript{g}$, as a function of an increasing electric field for over thirty nanoribbons of widths ranging from $n = 11$ ($w = 1.2$ nm) to $n = 43$ ($w = 5.1$ nm), grouped according to whether $n = 3p$, $3p+1$, or $3p+2$. Independently of the width, $\varepsilon\textsubscript{g}$ reduces and eventually closes at large enough ${E}\textsubscript{ext}$.  As the electrostatic potential between the two edges is proportional to their separation, $w$, the critical field required to develop the semimetallic phase, ${E}\textsubscript{crit}$, decreases as the nanoribbon widens \cite{Son2006b}. The dependence of the critical field on the width is shown in Fig.~\ref{Fig3}(d-f), where we observe that, within each class, ${E}\textsubscript{crit}$ is inversely proportional to the width of the nanoribbon,
\begin{equation}
E\textsubscript{crit} = \frac{V\textsubscript{crit}}{w} = \frac{2 V\textsubscript{crit}}{\sqrt{3}d\textsubscript{C}(n-1)},
\label{Eqn2}
\end{equation}
with $V\textsubscript{crit}$ being the critical potential required to close the energy gap. Fitting the first-principles results with Eqn.\ (2), we obtain $V\textsubscript{crit} = 22.45$, $18.69 $ and $10.87$~V for armchair graphene nanoribbons pertaining to the $n = 3p$, $3p+1$, and $3p+2$ classes, respectively. We thus envision that wider graphene nanoribbons belonging to the $3p+2$ class are optimally suited to electrically access Dirac fermions due to the occurrence of the semimetallic phase at the smallest fields.

\begin{figure}[t]
    \centering
    \includegraphics[width=1\columnwidth]{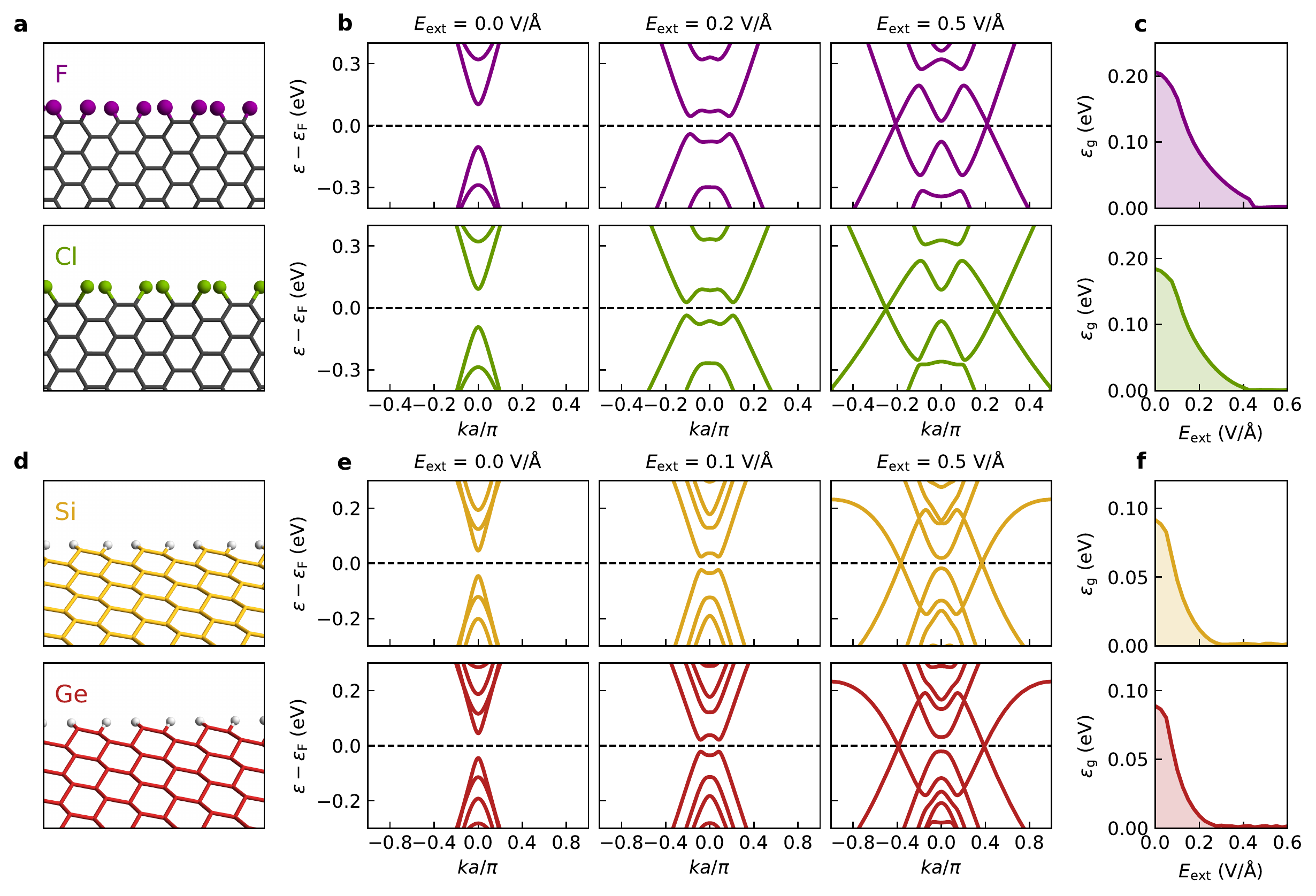}
    \caption{\textbf{Effect of the chemical composition of the nanoribbon on the field-induced Dirac fermions.} (a) Atomic structure of fluorine- and chlorine-terminated armchair graphene nanoribbons, (b) their band structures at width $n = 33$ in an external electric fields of strength ${E}\textsubscript{ext} = 0.0$, $0.2$, and $0.5$ V/{\AA}, along with (c) the evolution of their energy gap, $\varepsilon\textsubscript{g}$, with ${E}\textsubscript{ext}$. (d) Atomic structure of hydrogen-terminated armchair silicene and germanene nanoribbons, (e) their band structures at width $n = 33$ at fields of strength ${E}\textsubscript{ext} = 0.0, 0.1$, and $0.5$ V/{\AA}, and (f) the evolution of $\varepsilon\textsubscript{g}$ with ${E}\textsubscript{ext}$. Energies in the band structures are referenced to the Fermi level, $\varepsilon\textsubscript{F}$ \label{Fig4}}
\end{figure}

Our results are not restricted to hydrogen-terminated AGNRs. Motivated by the rapid experimental progress in the synthesis of halogen-terminated nanoribbons \cite{Panighel2020, Tan2013}, we assess the influence of fluorine and chlorine edge-passivations of AGNRs on the Dirac semimetallic phase, as shown in Fig.~\ref{Fig4}(a). The evolution of their band structure and energy gap with ${E}\textsubscript{ext}$ is presented in Fig.~4(b-c). No appreciable differences are observed as compared to their hydrogen-terminated counterparts, thereby suggesting that the type of edge termination does not affect the formation of the semimetallic phase. This is further supported by the results given in Supporting Figs.~S7-S8, where we explore additional monovalent terminations of varied electronegativities, including hydroxylated (-OH), amminated (-NH$_2$), and  hydroborated (-BH$_2$) nanoribbons. We therefore conclude that the field-assisted phase transition is robust with respect to the functional group used to passivate the carbon atoms at the edges.

To further illustrate that this field-induced transition to the Dirac semimetal is of a general nature, we investigate heavier group-IV congeners of graphene nanoribbons, \textit{i.e.}, silicene and germanene nanoribbons \cite{LeLay2009, DePadova2012a}. These nanostructures have been recently fabricated on silver substrates via epitaxial growth \cite{DePadova2008, DePadova2012b,Yuhara2021}, and subsequently integrated as active channels into electronic devices in various configurations \cite{Hiraoka2017, Tao2015}. Fig.~4(d) displays the atomic structure  of hydrogen-terminated silicene and germanene nanoribbons. Unlike carbon, the arrangement of silicon and germanium atoms in a honeycomb lattice is unstable in a planar geometry, and is subject to a structural distortion that consists of a relative out-of-plane displacement of the two sublattices by 0.53~{\AA} and 0.63~{\AA}, respectively \cite{Pizzochero2019, Cappelletti2021}. Despite the marked difference in the crystal structure, the field-induced semiconductor-to-semimetal transition is preserved, as evident from Fig.~\ref{Fig4}(e--f). The energy gap closes at weaker critical fields than in the case of graphene nanoribbons. The reason for this is that, for a given $n$, silicene and germanene nanoribbons are wider in $w$ than AGNRs owing to the longer bond distance ($d\textsubscript{Si} = 2.30$~{\AA} and $d\textsubscript{Ge} = 2.44$~{\AA} \emph{vs.} $d\textsubscript{C} = 1.42$~{\AA}). Altogether, these findings demonstrate a reversible electric control over Dirac fermions in this family of graphene-like nanoribbons regardless of their chemical composition, provided that the armchair edges of the honeycomb lattice and the ensuing $\pi$-electron network are retained.

\begin{figure}[t]
    \centering 
    \includegraphics[width=1\columnwidth]{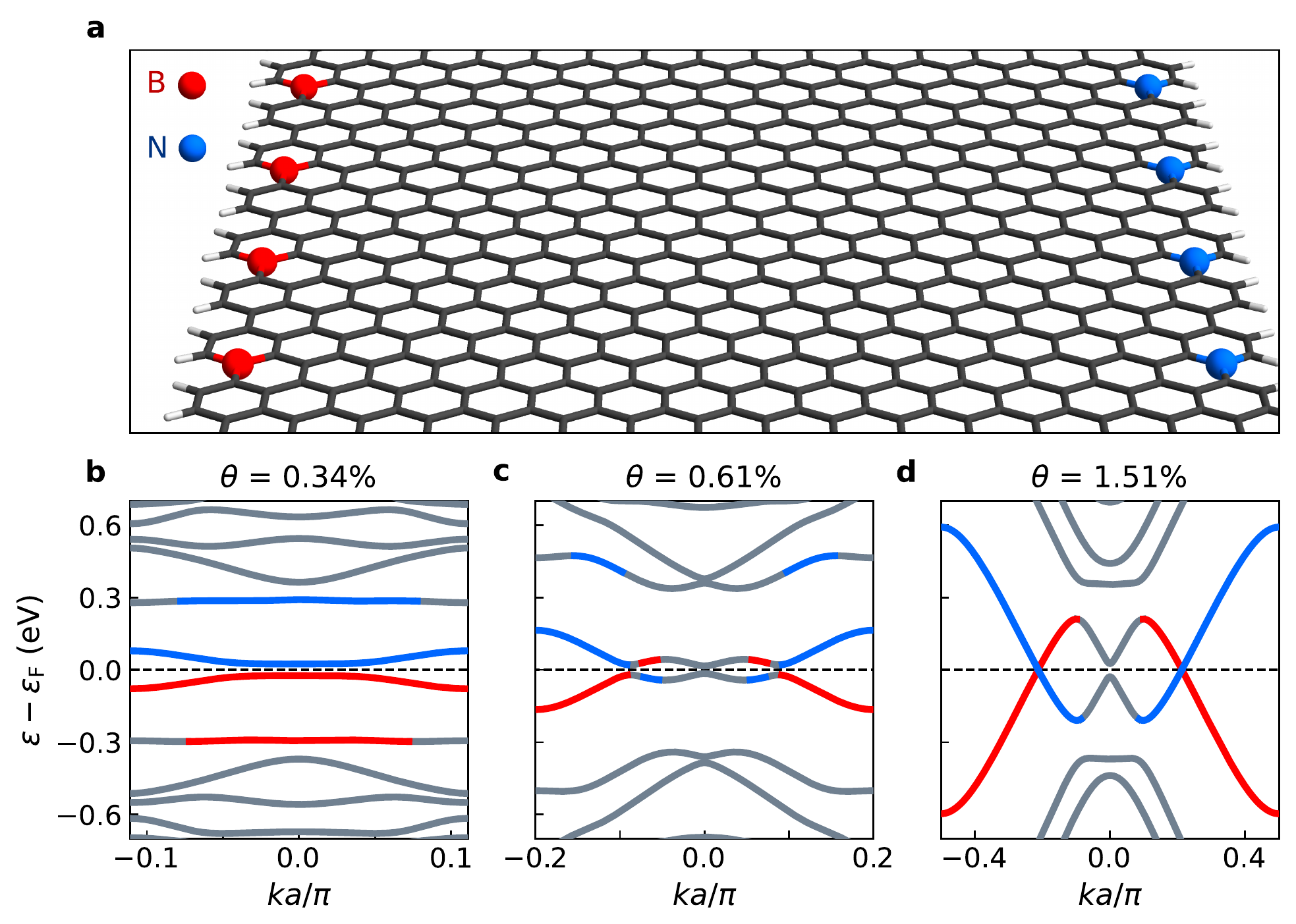}
    \caption{\textbf{Dirac fermions in graphene nanoribbons induced by chemical co-doping.} (a) Atomic structure of a hydrogen-terminated armchair graphene nanoribbon of width $n = 33$ patterned with an array of boron and nitrogen dopants, at a representative co-dopant density $\theta = 1.51$\%. (b-d) Band structures of co-doped AGNR at increasing $\theta = 0.34$, $0.61$, and $1.51$\%. Red, blue, and grey lines indicate the projections of the bands onto the B, N, and C atoms, respectively. Energies are referenced to the Fermi level, $\varepsilon\textsubscript{F}$. \label{Fig5}}
\end{figure}

\smallskip
\paragraph{Chemical route to Dirac fermions in AGNRs.} 
Finally, we propose an alternative, chemical approach for creating Dirac fermions in AGNRs. Instead of relying on an external electric field, we consider the incorporation of $n$-type impurities into one edge of the nanoribbon and $p$-type impurities into the other, effectively leading to an intra-ribbon $p$-$n$ junction and an accompanying transverse internal field. This qualitatively corresponds to the physical realization of the simple tight-binding model explored above, where edges of opposite polarities were considered; see Fig.\ \ref{Fig2}(a,b) and related discussion. In Fig.~\ref{Fig5}(a), we show the atomic structure of an AGNR with $n = 33$ ($w = 3.9$ nm) patterned with two arrays of boron and nitrogen dopants. We remark that advances in the on-surface synthesis enable atomically precise incorporations of heteroatoms in graphene nanoribbons \cite{Yano2020}, including periodically arranged nitrogen \cite{Cai2014, Zhang2014}  or boron atoms, \cite{Kawai2015, Cloke2015} as well as their co-doping \cite{Kawai2018}.

For a given width of the nanoribbon, the strength of the internal field resulting from the spatial separation of electrons and holes can be quantified through the density of the chemical dopants hosted in the lattice, $\theta = \left[n\textsubscript{X} /   (n\textsubscript{X} +  n\textsubscript{C})\right] \%$, where $n\textsubscript{X}$ and $n\textsubscript{C}$ are the total numbers of dopant and carbon atoms in the nanoribbon. In Fig.~\ref{Fig5}(b-d), we show the band structure of a hydrogen-terminated AGNR co-doped with boron and nitrogen atoms at $\theta = 0.33$, $0.61$ and $1.51$\%. Upon ambipolar co-doping, we observe a rearrangement of the bands around the Fermi level, similar to that induced through the application of the external electric field; \emph{cf.}\ Fig.~\ref{Fig1}(b). The Dirac semimetallic phase emerges at dopant densities exceeding $\sim$$1.5$\%, with the linearly crossing bands mainly arising from the heteroatoms. 
 
\smallskip
\paragraph{Summary and conclusions.} We have demonstrated that semiconducting armchair graphene nanoribbons develop a Dirac semimetallic phase under a transverse electric field. Such a field can be achieved by laterally gating the nanoribbon or by introducing  $n$- and $p$-type dopants along the nanoribbon edges. The semimetallic phase features linearly dispersing bands crossing at the Fermi level and hosts zero-energy Dirac fermions propagating along the nanoribbon edges. The semiconductor-to-semimetal transition occurs at critical fields that scale inversely with the nanoribbon width and is universal to group-IV armchair nanoribbons, including silicene and germanene nanoribbons, irrespective of the edge passivation. Our work opens new avenues to electrically engineer Dirac fermions in this broad family of one-dimensional nanostructures.

\smallskip
\paragraph{Acknowledgments.}
M.P.~is grateful to QuanSheng Wu (EPFL), Rocco Martinazzo (University of Milan), and Nick R.~Papior (Technical University of Denmark) for fruitful interactions. M.~P.~is supported by the Swiss National Science Foundation (SNSF) through the Early Postdoc.Mobility program (Grant No.~P2ELP2-191706) and the NSF DMREF (Grant No. 1922165). N.~V.~T. is supported by the President's PhD Scholarship of Imperial College London.

\bibliography{References}

\end{document}